**Electronic structure and magnetic properties of graphene/Co composites**

I.S. Zhidkov[1,2], N.A. Skorikov[1*], A.V. Korolev[1], A.I. Kukharenko[1,2], E.Z. Kurmaev[1],

V.E. Fedorov[3] and S.O. Cholakh[2]

[1]*Institute of Metal Physics, Russian Academy of Sciences-Ural*

*S. Kovalevskoi 18 str., 620990 Yekaterinburg, Russia*

[2]*Ural Federal University, 19 Mira Str., 620002 Yekaterinburg, Russia*

[3]*Nikolaev Institute of Inorganic Chemistry, Siberian Branch of Russian Academy of Sciences,*

Abstract. The results of measurements of XPS spectra and magnetic properties of graphene/Co composites prepared by adding of $CoCl_2 \cdot 6H_2O$ diluted in ethyl alcohol to highly-splitted graphite are presented. XPS Co *2p* measurements show two sets of $2p_{3/2,1/2}$-lines belonging to oxidized and metallic Co-atoms. This means that metal in composite is partly oxidized. This conclusion is confirmed by magnetic measurements which show the presence of the main (from the metal) and shifted (from the oxide) hysteresis loops. The presence of oxide layer on the metal surface prevents the metal from the full oxidation and (as shown by magnetic measurements) provides the preservation of ferromagnetic properties after long exposure in ambient air which enables the use of graphene/metal composites in spintronics devices.

## 1. Introduction

The structural resemblance of graphene and close-packed *3d*-metals: Ni(111) and Co(111) with a lattice mismatch of about 1% induces the studies of graphene-metal contacts and interfaces [1-7]. There is also the growing theoretical and experimental interests in the use of graphene in magnetic sandwich structures in the search for new materials for magnetic tunnel junction (MTJ) applications [8-11]. It is found that graphene/ferromagnetic metal interfaces play an important role in graphene-based spintronic devices and spin injection and spin transport have been realized at room temperature in graphene-based spin valves [12]. The magnetic properties of ferromagnet/graphene interfaces were studied both experimentally and theoretically [6, 13-14]. The using graphene in spin transport and injection require a detailed understanding of the electronic structure and magnetic properties of the graphene/ferromagnet interfaces and composites. In connection with this in the present paper we have performed XPS and magnetic measurements of graphene/Co composites. The results of XPS measurements of the valence band spectra are compared with specially performed density functional calculations of electronic structure of graphene/Co/CoO interfaces.

## 2. Experimental and Calculation Details

The highly-splitted graphite with very developed surface (300 $m^2$/g) obtained by rapidly heating of the intercalated graphite compound $C_2F(BrF_3)_{0.12}$ to 800 ºC (regime of «thermal shock») was used as carbon matrix. The transmission electron microscopy (TEM) measurements show (Fig. 1) that obtained by such a way a final product consists of thin graphene layers with thickness of 2-7 nm which can be related low-layered graphene.

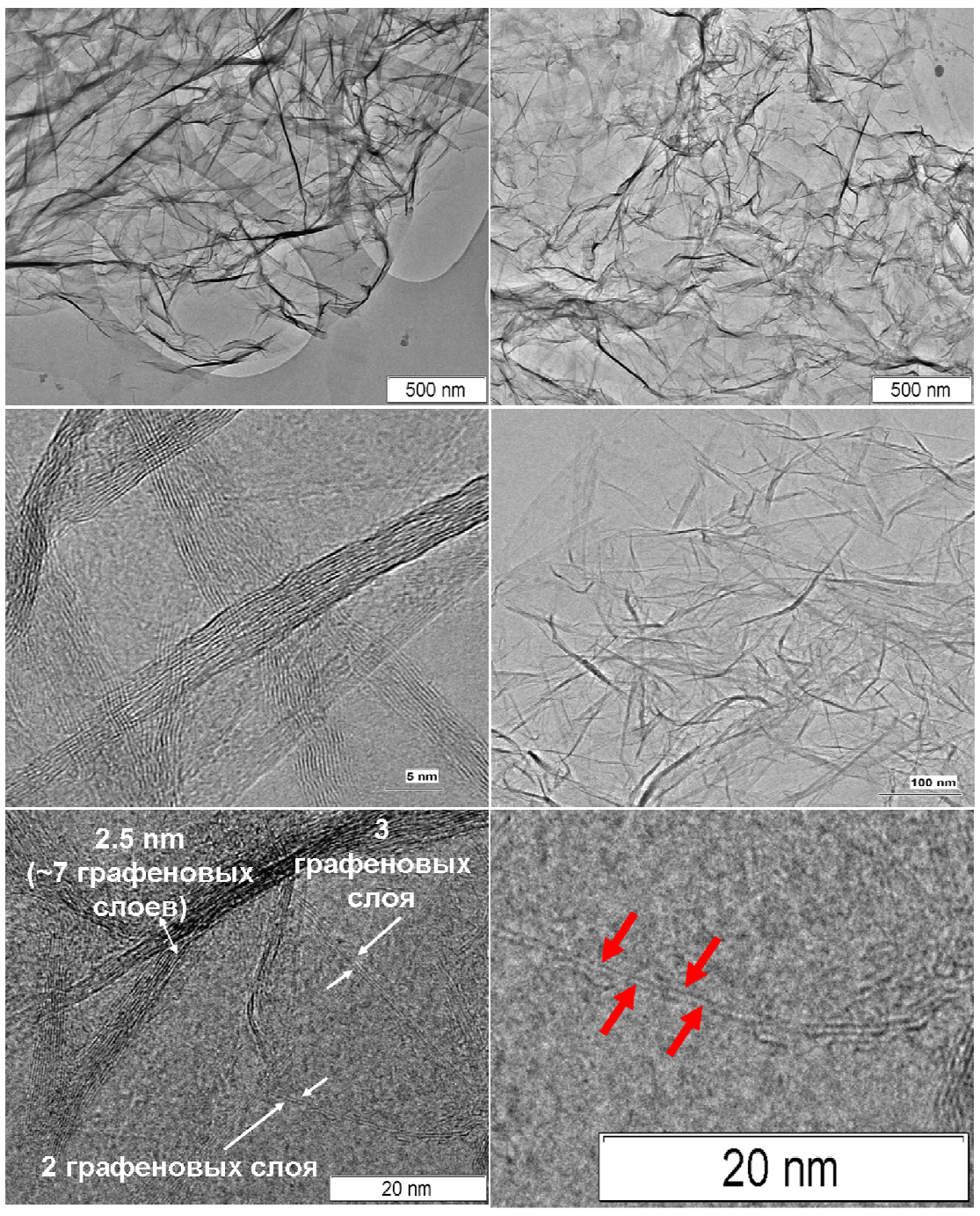

Fig. 1. TEM images for different areas of composites, obtained by thermal decomposition of $C_2F(BrF_3)_{0.12}$ intercalate

The graphene/Co composites were prepared by adding of 10 mg of $CoCl_2 \cdot 6H_2O$ diluted in ethyl alcohol to 100 mg of carbon matrix. Then the mixture was stirred, dried in air during 8 hours, pressed into pellets (dia of ~12 mm) and annealed in quartz tube at 400 $^0$C for 2 hours. Afterwards the synthesized composites were cooled in hydrogen flow up to room temperature.

The XPS spectra were recorded using Al Kα x-ray emission; the spot size was 100 μm, and the x-ray power load on the sample was kept below 25 watts. Typical signal to noise ratios were above 10000:3. The spectra were processed using ULVAC-PHI MultiPak Software 9.3. XPS spectra were calibrated using a reference energy of 285.0 eV for the carbon *1s* level.

The SQUID magnetometer MPMS-5XL (Quantum Design) was used for measuring magnetization curves $M(H)_{T=const}$ ($T$ is a temperature) in magnetic fields $H$ up to 50 kOe and temperature dependences of magnetization $M(T)_{H=const}$ in the temperature range 2–400 K. Magnetic AC-susceptibility was measured in an alternating magnetic field $h = h_A sin(2\pi f)$ at $h_A$ = 4 Oe and $f$ = 80 Hz.

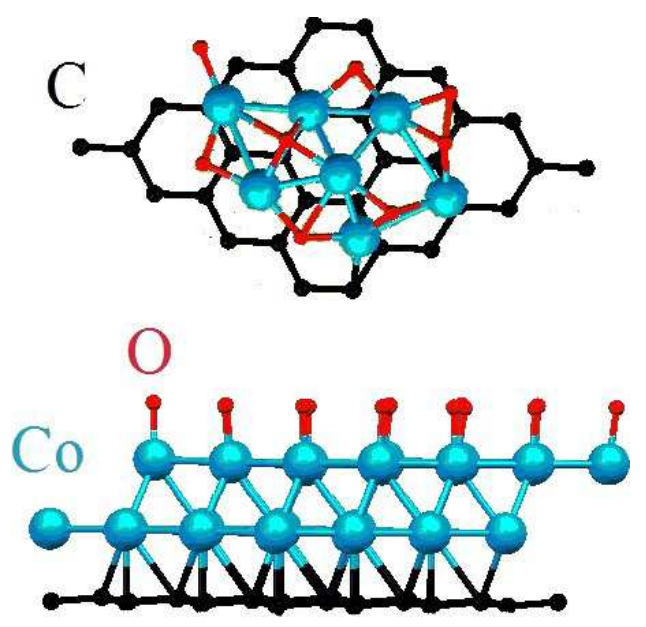

Fig. 2. The optimized atomic structure for Gr/Co/CoO composite .

The pseudo-potential Quantum-Espresso code [15] was used for the investigation of magnetic and spectral properties of Graphene/CoO/Co composite. We used ultrasoft pseudo-potential with Perdew-Wang (PW) version of exchange potential [16]. For the simulation of the Graphene/CoO/Co composite the following model was used: one layer of graphene was covered by one (for Graphene/CoO) or two (for Graphene/CoO/Co) layers of *fcc* (111) cobalt and than one layer of oxygen atoms was placed. On the top of oxygen layer it was added 10 Å of empty space. The atomic positions in the constructed cells were relaxed while each component of force were more than 2 mRy/a.u. Relaxation and final calculations were performed with a k-point mesh of 8x8x1 in the Monkhorst-Pack scheme [17]. The plane-wave and kinetic energy cut-off's were chosen to be 45 Ry and 200 Ry, respectively.

## 3. Results and Discussion

XPS survey spectrum of graphene/Co composite is presented in Fig. 3 (upper panel). One can see the strong C *1s* signal and weak Co *2p* and O *1s* lines; XPS survey spectrum does not show the presence of any additional impurities.

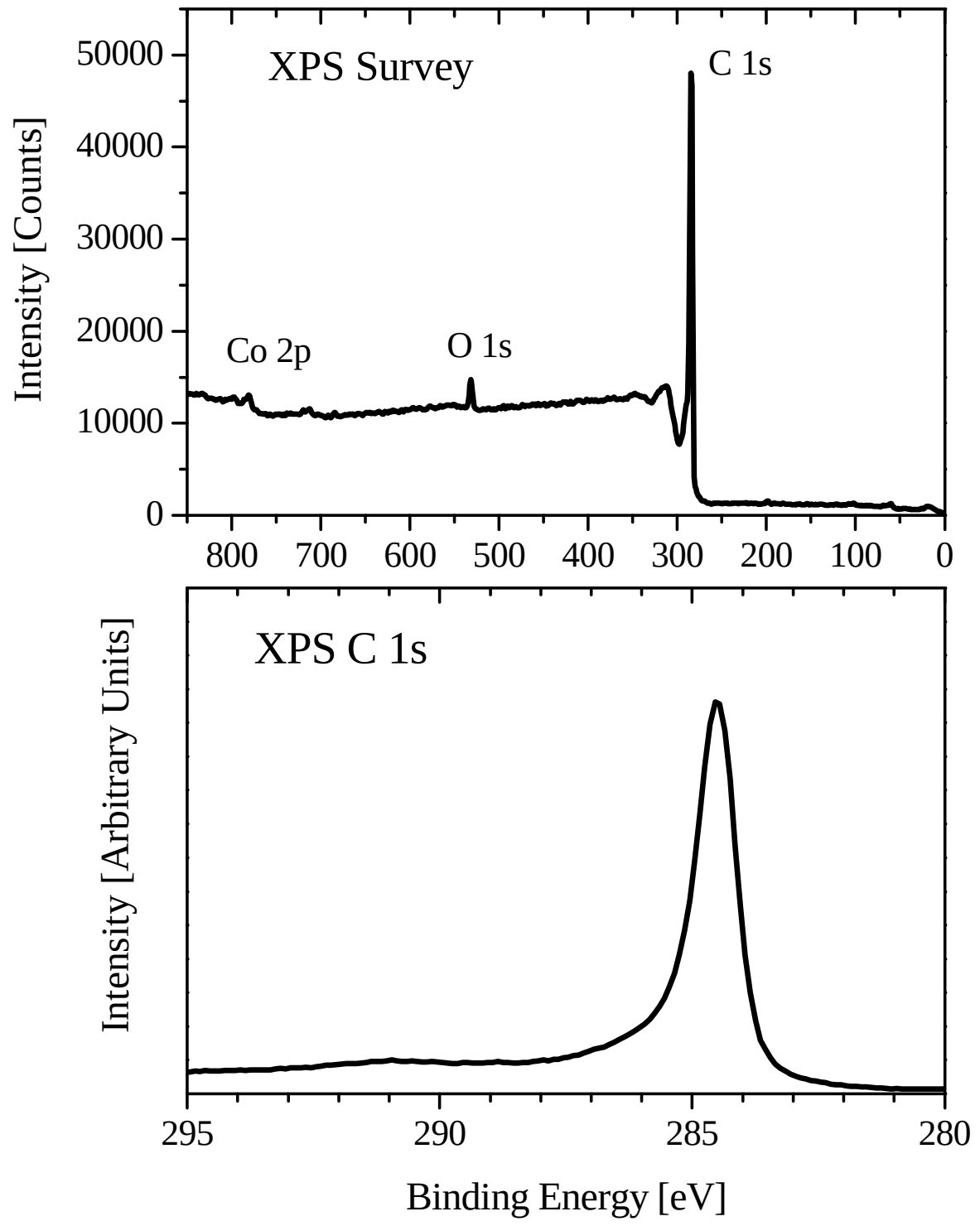

Fig. 3. Upper panel: XPS survey spectrum of Graphene/Co composite.

Lower panel: High-energy resolved XPS C *1s* of Graphene/Co composite.

The high-energy resolved XPS C*1s* spectrum of Graphene/Co composite (Fig. 3, lower panel) is found to be identical to that of graphene [18] without any traces of C-O-C, C-OH and O=C-O-O contributions.

The chemical state of Co atoms in Graphene/Co composite can be determined with help of measurements of XPS Co $2p_{3/2.1/2}$-core levels. As seen from the Fig. 4, the high-energy resolved XPS Co $2p$-spectrum of Graphene/Co show two sets of $2p_{3/2,1/2}$-lines which can be attributed to superposition of $Co^0$ and $Co^{2+}$ signals.

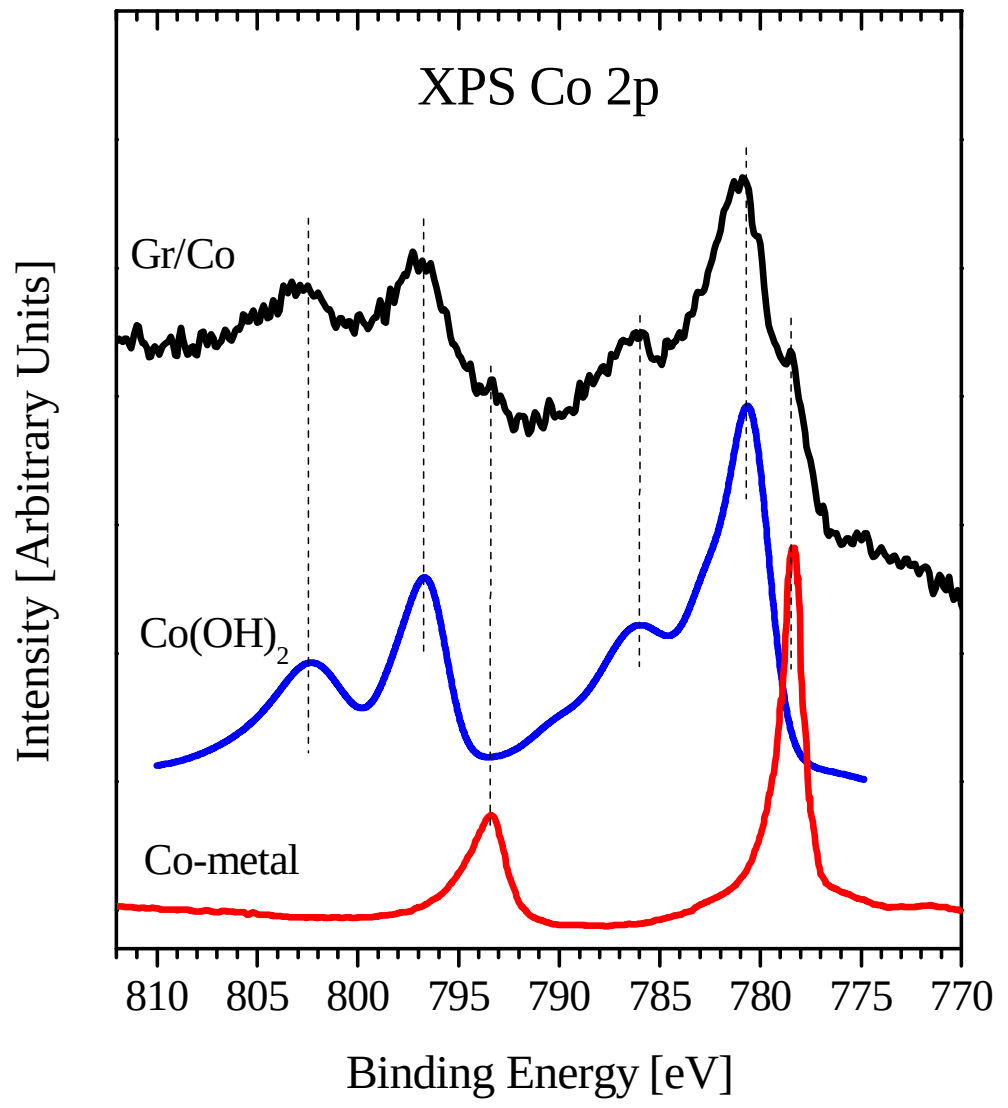

Fig. 4. The comparison of XPS Co *2p* of Graphene/Co composite with spectra of reference samples (Co-metal and $Co(OH)_2$ [19]).

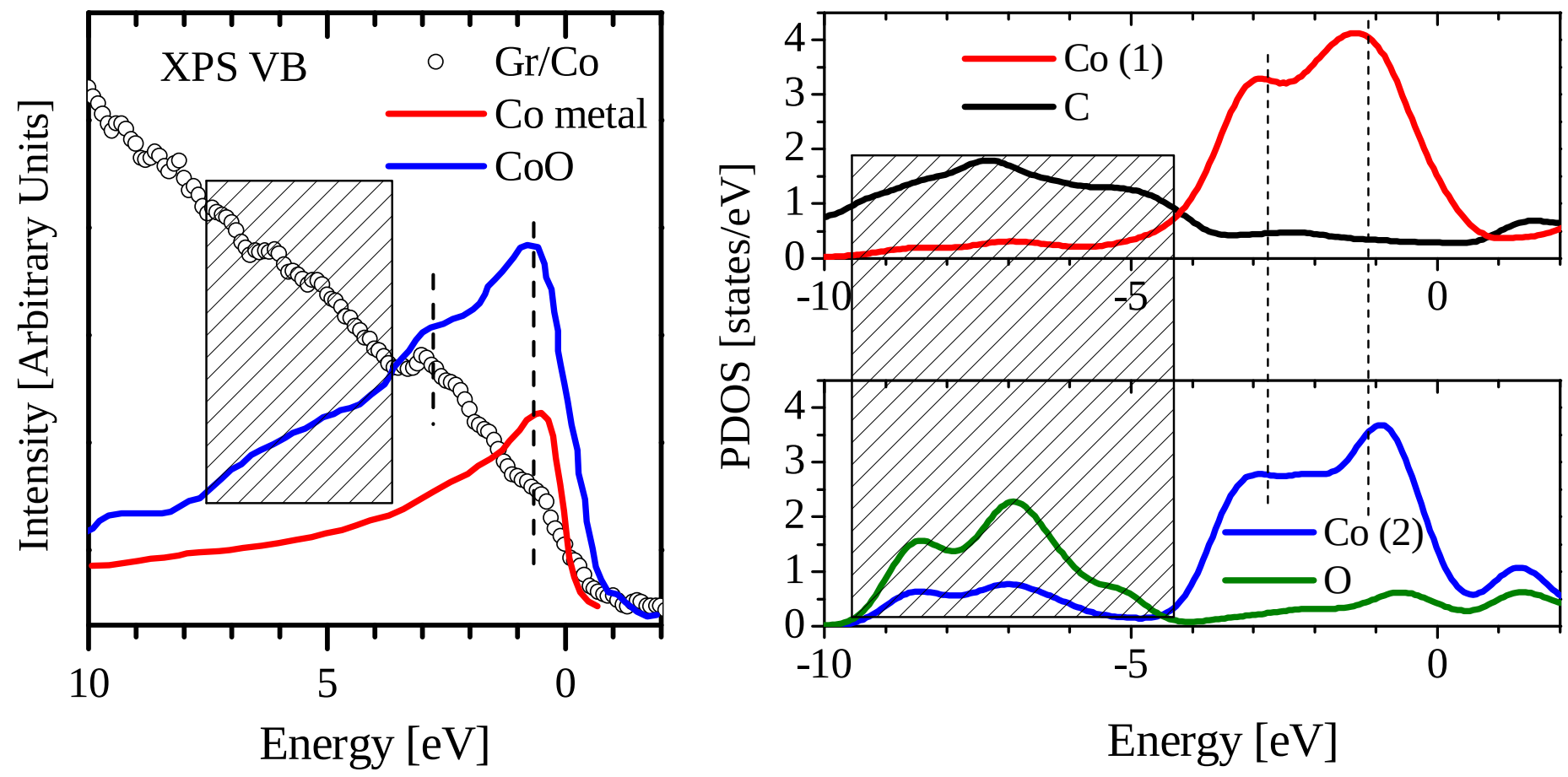

Fig. 5. Left panel: Comparison of XPS VB spectra of Graphene/Co with those of Co [20] and CoO [21]; Right panel: Calculated partial densities of states for Graphene/Co/CoO composite.

This conclusion is confirmed by the of XPS valence band spectrum of Graphene/Co composite showing a rather complicated structure (Fig. 5, left panel) which can be related to contributions of graphene, Co metal and CoO. The results of XPS VB measurements are in a good agreement with density functional theory calculations of the electronic structure of Graphene/Co/CoO composite (Fig. 5, right panel). As seen, the contribution of partial density of states (PDOS) of Co-metal atoms (Co1) dominate near the Fermi level whereas contributions of PDOS of Co(2) and O atoms from the upper CoO layer are located at the lower part of the valence band. The spin magnetic moments of Co ions are found to be ferromagnetically arranged in the case of two layers of cobalt. The magnetic moments values are found to be different: 0.76 $\mu_B$/atom for oxidized layer and 0.96 $\mu_B$/atom for the layer closest to the graphene. For cell containing one layer of Co ions we have received nonmagnetic solution.

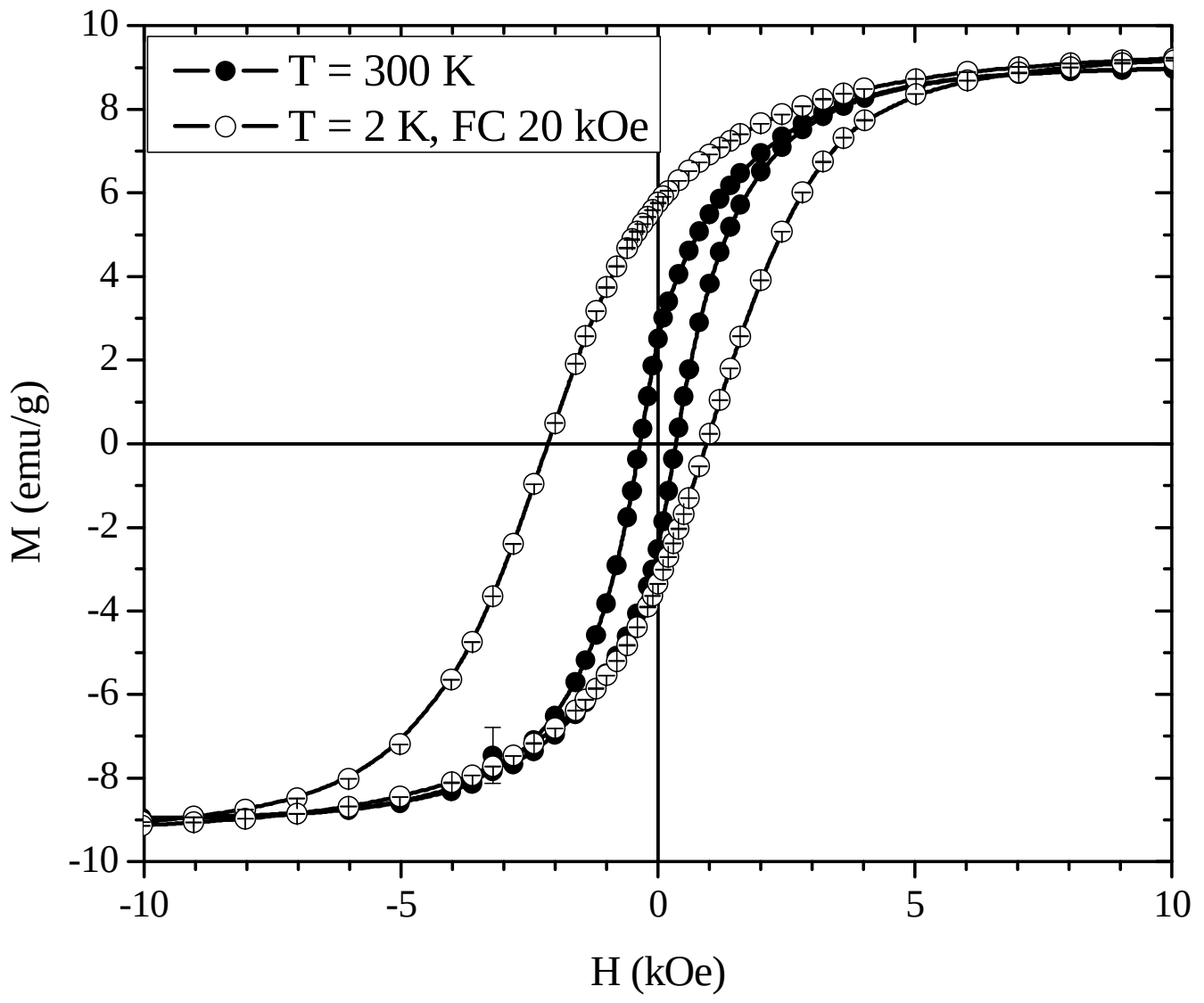

Fig. 8. The hysteresis loops for Graphene/Co measured at T=300 K and T=2 K after cooling in H=20 KOe.

The results of DFT calculations of magnetic properties are confirmed by magnetic measurements of Gr/Co/CoO composites after 6 months exposure in ambient air. The hysteresis loops of Gr/Co (m=20,74mg) were measured at T=300 K and T=2 K after cooling in H=20 kOe (FC regime). As seen from Fig. 8, the strong effect of hysteresis loop shift is observed at T=2 K. This effect most probably is due to the fact that Co particles contain of antiferromagnetic cobalt oxide (CoO). The temperature dependence of the real ($\chi'$) and imaginary ($\chi''$) components of the dynamic AC-susceptibility for Graphene/Co is shown in Fig. 9. The low-temperature feature of $\chi'(T)$ function (see inset) apparently is due to small amount of antiferromagnetic type oxide (CoO) for presence of which the shifted hysteresis loop is indicated (see Fig. 8).

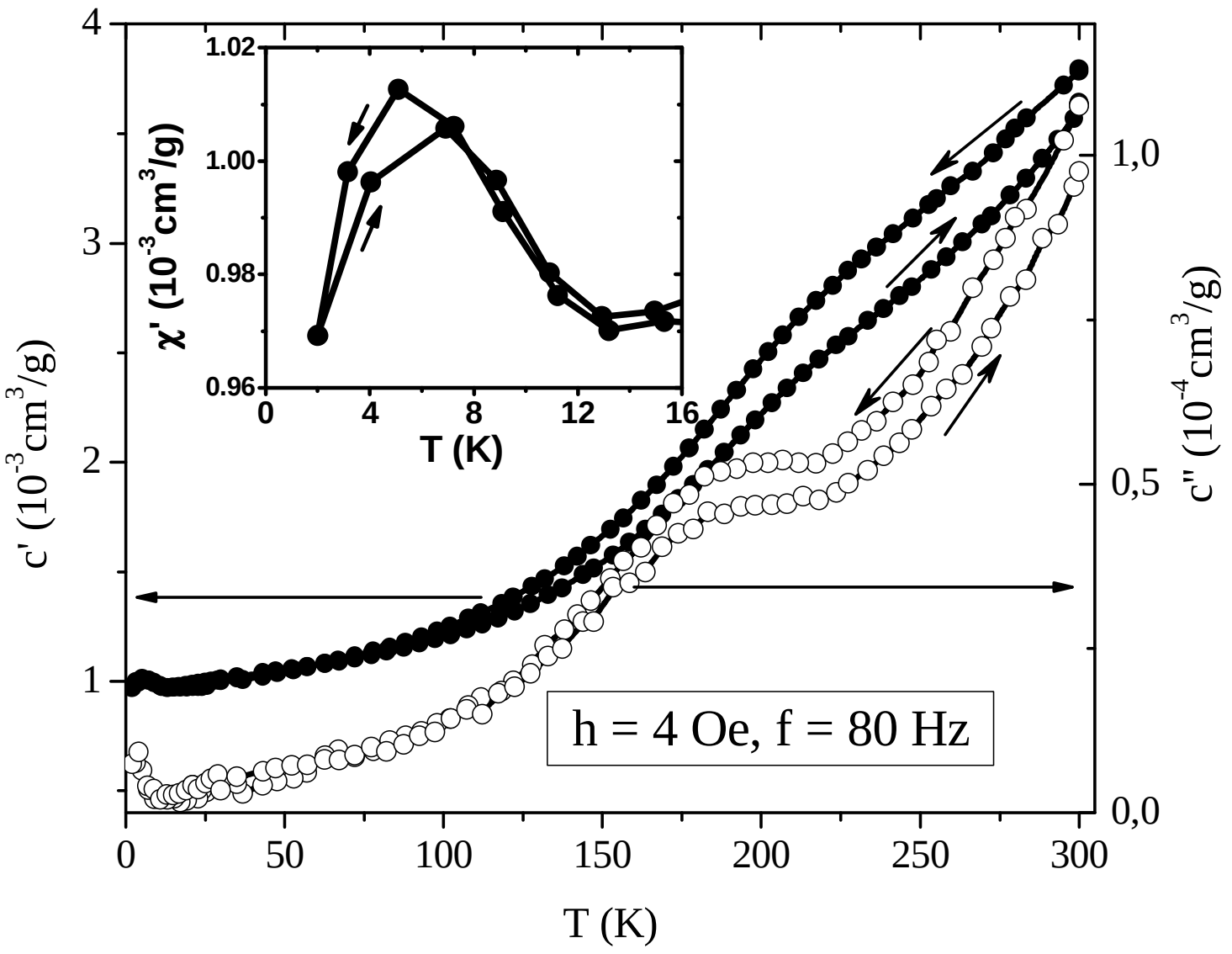

Fig. 9. The temperature dependence of magnetic AC-susceptibility of Graphene/Co.

**Conclusion**

Basing on results of DFT calculations and XPS measurements one can conclude that Co atoms are partly oxidized in Graphene/Co composite. For two or more layers of Co atoms located beneath the oxidized CoO layer the ferromagnetic behavior is theoretically predicted which is confirmed by magnetic measurements. Therefore the formation of protective oxide layer prevents Co from deep oxidation and allows to use its ferromagnetic properties in spintronics devices.

**Acknowledgements**

The results of experimental measurements of XPS spectra and calculations of electronic structure were supported by the grant of the Russian Scientific Foundation (project no. 14-22-00004).

*Corresponding author, E-mail: skorikov@ifmlrs.uran.ru

References:

1. Tesfaye Abtew, Bi-Ching Shih, Sarbajit Banerjee and Peihong Zhang, Nanoscale, 5 (2013)

2. G. Bertoni, L. Calmels, A. Altibelli and V. Serin, Phys. Rev. B 71 (2005) 075402.

3 P. A. Khomyakov, G. Giovannetti, P. C. Rusu, G. Brocks, J. van den Brink and P. J. Kelly, Phys. Rev. B 79 (2009) 195425.

4. Y. S. Dedkov, M. Fonin and C. Laubschat, Appl. Phys. Lett. 92 (2008) 052506.

5. Y. Murata, V. Petrova, B. B. Kappes, A. Ebnonnasir, I. Petrov, Y.-H. Xie, C. V. Ciobanu and S. Kodambaka, ACS Nano, 4 (2010) 6509.

6. Y. S. Dedkov and M. Fonin, New J. Phys. 12 (2010) 125004.

7. M. Fuentes-Cabrera, M. I. Baskes, A. V. Melechko and M. L. Simpson, Phys. Rev. B 77 (2008) 035405.

8. E. Cobas, A. L. Friedman, O. M. J. van't Erve, J. T. Robinson and B. T. Jonker, Nano Lett. 12 (2012) 3000.

9. T. Mohiuddin, E. Hill, D. Elias, A. Zhukov, K. Novoselov and A. Geim, IEEE Trans. Magn.44 (2008) 2624.

10. R. Sato, T. Hiraiwa, J. Inoue, S. Honda and H. Itoh, Phys. Rev. B 85 (2012) 094420.

11. V. M. Karpan, G. Giovannetti, P. A. Khomyakov, M. Talanana, A. A. Starikov, M. Zwierzycki, J. van den Brink, G. Brocks and P. J. Kelly, Phys. Rev. Lett. 99 (2007) 176602.

12. N. Tombros, C. Jozsa, M. Popinciuc, H. T. Jonkman, and B. J. Van Wees, Nature (London) 448 (2007) 571.

13. M. Weser, Y. Rehder, K. Horn, M. Sicot, M. Fonin, A. B. Preobrajenski, E. N. Voloshina, E. Goering, and Yu. S. Dedkov, Appl. Phys. Lett. 96 (2010) 012504.

14. G. Bertoni, L. Calmels, A. Altibelli, and V. Serin, Phys. Rev. B 71 (2005) 075402.

15. P. Giannozzi, S. Baroni, N. Bonini, N., et al., J. Phys. :Condens. Matter 21 (2009) 395502.

16. J.P. Perdew, A. Zunger, Phys. Rev. B 23 (1981) 5048.

17. H.J. Monkhorst, J.D. Pack, Phys. Rev. B 13 (1976) 5188.

18. M. J. Webb, P. Palmgren, P. Pal, O. Karis, and H. Grennberg, arXiv:cond/mat 1103.6177.

19. Mark C. Biesinger, Brad P. Payne, Andrew P. Grosvenor, Leo W.M. Lau, Andrea R. Gerson, Roger St.C. Smart, Appl. Surf. Sci. 257 (2011) 2717.

20. Timo Hofmann, Ted H. Yu, Michael Folse, Lothar Weinhardt, Marcus Bar, Yufeng Zhang, Boris V. Merinov, Deborah J. Myers, William A. Goddard, III, and Clemens Heske, J. Phys. Chem. C 116 (2012) 24016.

21. S. Uhlenbrock, Ph.D. thesis, Fachbereich Physik der Universität, Osnabrück, 1994.